# *Quantum-Gravity Phenomenology and the DSR Ether Theories.*

*Kenneth M. Sasaki.*

**Abstract.**

Guided primarily by versions of a theoretical framework called Doubly Special Relativity, or DSR, that are supposed to entail speeds of light that vary with energy while preserving the relativity of inertial frames, quantum-gravity phenomenologists have recently been seeking clues to quantum gravity, in hoped-for differing times of arrival, for light of differing energies, from cosmologically distant sources. However, it has long been known that signals, of arbitrarily high speed in opposing directions, could be used to observe the translational state of (absolute) rest, as could signals of a fixed speed different from *c*. Consequently, the above versions of DSR are nonviable.

## I. Introduction.

Recently, phenomenologists have been seeking clues to quantum gravity, in hoped-for differing times of arrival, for light of differing energies, from cosmologically distant sources (see [1] and its references). The primary guiding theoretical framework is a class of models intended to modify Special Relativity, called Doubly Special Relativity (DSR) [2-8,1], having dispersion relations deformed in a way thought to preserve relativity, with respect to both an approximate low-energy light speed and a minimum observable length. While some versions of DSR are supposed to leave the speed of light unchanged, versions which we will call VLS-DSR, espoused by the *variable-light-speed* movement, have speeds of light that vary with increasing energy, either faster or slower than *c* (see [1] and its references for further information about the various DSR versions).

However (as we will further discuss), arbitrarily fast signals, in opposing directions, could be used to observe true simultaneity [9a,10a,11] and thus the unique translational state of "rest" – "the ether" – only in which time runs fastest and only with respect to which the speed of heretofore-observed light is the same in all directions [11].[1]

Furthermore, just over one hundred years ago, Poincaré wrote (in translation) [13], "What would happen if one could communicate by non-luminous signals whose velocity of propagation differed from that of light? If, after having adjusted the watches by the optical procedure, we wished to verify the adjustment by the aid of these new signals, we should observe discrepancies which would render evident the common translation of the two stations." (Poincaré had in mind "non-luminous signals" that, like light, would always travel at a fixed speed.)

What Poincaré essentially realized (and we will also further discuss) is that the ether is indistinguishable in trivial flat space-time, with just ordinary light traveling at fixed speed, because time dilation, length contraction, and the increase of mass with motion occur just so, to give the illusion that these effects occur identically with respect to all attainable states of motion. But these effects could not occur just so, with two types of signals that always respectively travel at two differing fixed speeds; so such signals would make the ether observable. (Sjödin has also noted that two such differing "wave"-type signals could be used to determine rest [11].)

---

[1] What we here call "rest", or "the ether", is sometimes called "absolute rest", but the present author does not like this, since the dragging of inertial frames makes the "absolute" dubious [12].

We will respectively apply and adapt the above two methods of observing the ether, to VLS-DSR's variable speeds of light, demonstrating that VLS-DSR would make the ether observable. Since VLS-DSR purports to preserve relativity, it is thus nonviable.

Section II reviews important aspects of clock synchronization. Section III reviews how VLS-DSR light with diverging speeds could be used to observe the ether. Section IV presents a procedure by which VLS-DSR light, either faster or slower than $c$, could be used to achieve rest. Section V contains general comments and conclusions. And Sections VI and VII are respectively acknowledgements and references.

## II. Clock Synchronization in Trivial Flat Space-time.[2]

Suppose two clocks, *A* and *B*, to comove in a trivial flat space-time. An observer at *A* sends an ordinary low-energy light signal towards *B*, when *A* reads *time $t_1$*; an observer at *B* receives the signal and reflects it back, when *B* reads $t_2$; and the observer at *A* receives the signal back, when *A* reads $t_3$. Poincaré (essentially) [14] and later Einstein [15] *defined A* and *B* to be synchronized if:

1) $$t_2 - t_1 = t_3 - t_2.$$

However, as Poincaré realized, with previously observed light as the fastest signal and only fixed-speed propagator, there is a dilemma [13]: As Reichenbach put it [10b], "To determine the simultaneity of distant events we need to know a velocity, and to measure a velocity we require knowledge of the simultaneity of distant events." So, while our synchronization procedure measures $t_3 - t_1$ at *A*, $t_2$ at *B* is only measured to have an exact value in the future of $t_1$ and the past of $t_3$ [9b,10b]. Reichenbach accounted for the $t_2$ possibilities and associated light-velocity combinations, by letting $0 < \varepsilon < 1$ and defining [9b,10b]:

2) $$t_2 = t_1 + \varepsilon(t_3 - t_1).$$

Figure 1) shows velocities, between our clocks and light, that are consistent with our synchronization-procedure measurements, with both light propagating symmetrically and our clocks at rest $\Leftrightarrow \varepsilon = 1/2$ {reducing Equation 2) to Equation 1) [9c,10b]}.

Following convention, we will refer to as "standard" the *assumption* that $\varepsilon = 1/2$, and for each state of motion, so also name the resulting synchrony and reference frame.

There is a one-to-one correspondence between the *actual* values of $\varepsilon$ and the lines of synchrony outside any ordinary low-energy light cone, with one line of any given synchrony running through points of $((t_3 + t_1)/2, A)$ and $(t_2, B)$. Figure 2) shows two lines of synchrony for $\varepsilon = 3/4$, as seen in the standard frame for $\varepsilon = 1/2$. Furthermore, each synchrony is standard for only one state of motion, and each state of motion has only one standard synchrony. Composing creates a one-to-one correspondence, between the actual values of $\varepsilon$ and the states of motion within any ordinary low-energy light cone. For each such state of motion, the actual value of $\varepsilon$ corresponds to the line of standard synchrony, produced by the standard assumption for $\varepsilon$, with the actual and assumed $\varepsilon$ values equal (to 1/2) only for rest. The more the actual value of $\varepsilon$ differs from 1/2, the more the corresponding state of motion differs from rest. In Figure 1B), if the clocks in the state of $\varepsilon = 3/4$ were synchronized using our above procedure, under the standard assumption that they were in the state of $\varepsilon = 1/2$, they would have the standard synchrony of $\varepsilon = 3/4$, with the clock world lines and synchrony lines defining the standard frame of $\varepsilon = 3/4$.

---
[2] Much of this section was adapted from [12], with permission of the author.

Figure 1)

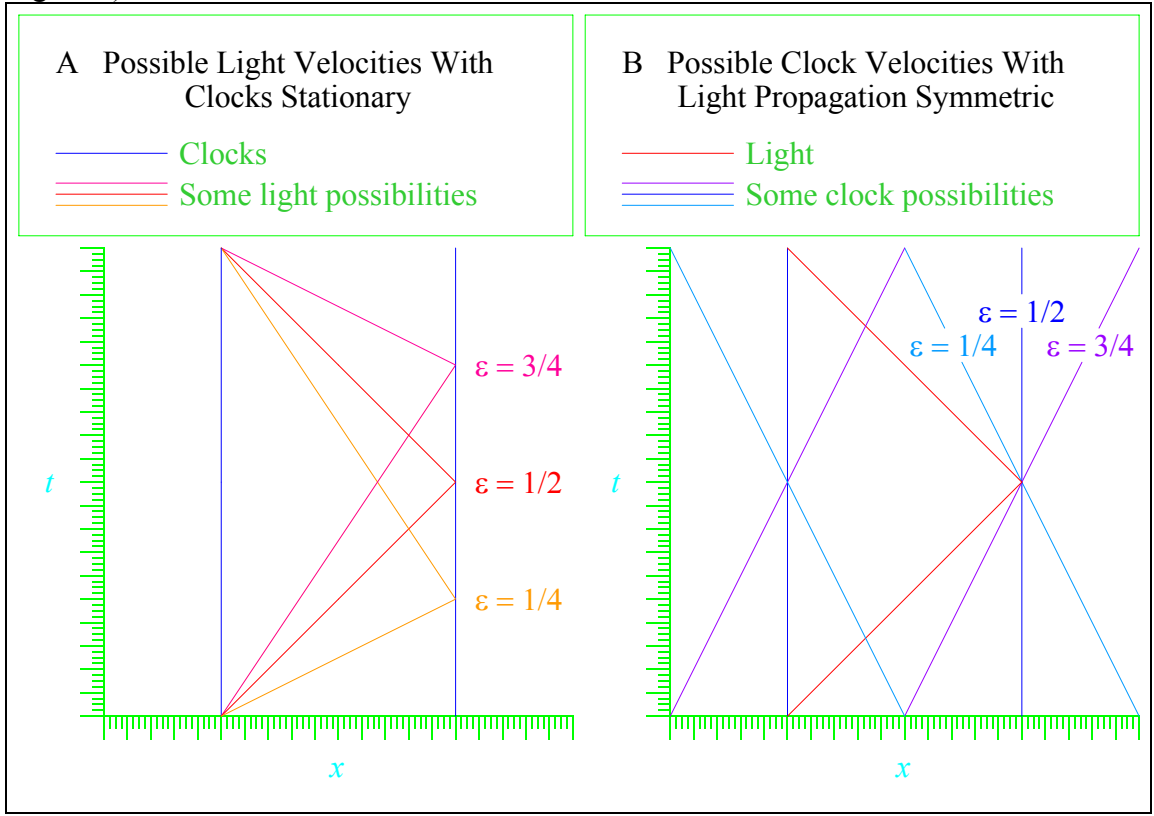

Figure 2)

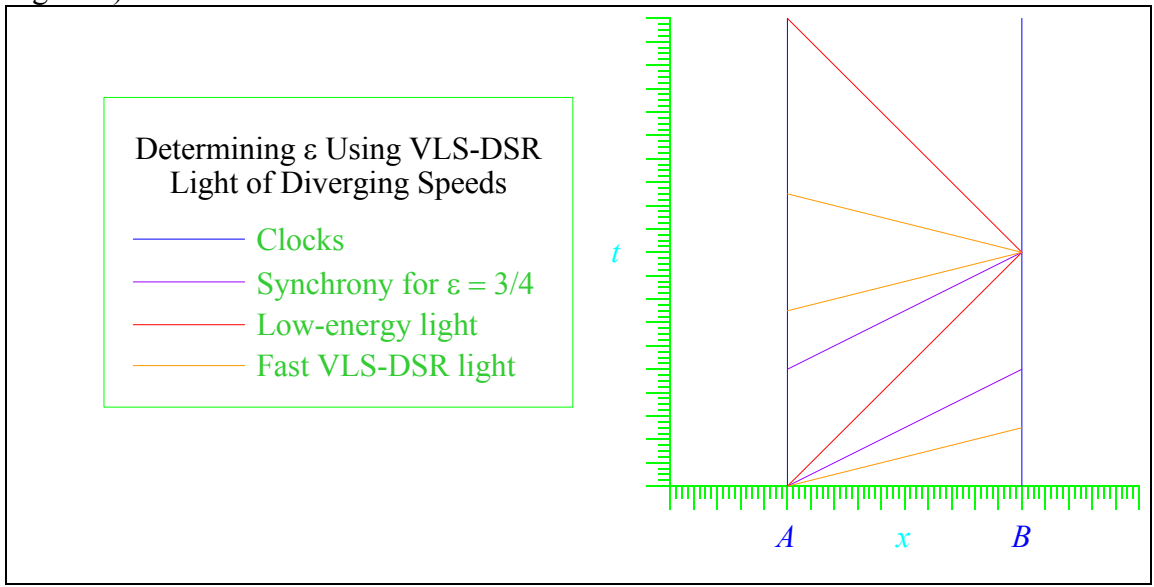

### III. Observing the Ether, Using VLS-DSR Light of Diverging Speeds.

With relativity, signals of any speed exceeding $c$, in all directions of all standard frames, would imply infinitely fast signals, and even signals that propagate backwards in time – with negative speeds – thus allowing causal paradoxes [16].

However, arbitrarily fast signals, propagating with respect to the ether (a single state of motion), would cause no paradox [11] or even the appearance of one, since in any frame, any chain of such signals, that would begin and end at the same spatial coordinates, would end at least slightly time-forward from its beginning. Any one-way signal, faster than ordinary light, would appear to propagate backwards in time, in some standard frames; but such appearances would be illusions of misunderstood synchrony.

Avoiding causal paradox is reason enough to conclude that any arbitrarily fast propagation would be with respect to the ether. But furthermore, light velocity diverging with energy leaves no room in the energy spectrum for light not propagating forwards in time. A flat world line would require infinite energy which could never actually be achieved, so any frame showing such must be illusory, again demanding the ether.

Now, consider fast VLS-DSR light signals, like those in Figure 2), for which $t'_1$, $t'_2$, and $t'_3$ are analogous to $t_1$, $t_2$, and $t_3$, with which we would have:

3) $$t'_3 - t'_1 < t_3 - t_1.$$

With such signals, observation could decrease the indeterminacy of $\varepsilon$, to an interval smaller than (0,1), in proportion to Inequality 3) [9a,10a] {keeping (0,1) defined by ordinary low-energy light [17]}. Precisely those states of motion, including rest, with lines of standard synchrony outside the fast VLS-DSR light cone, would be associated with values inside the smaller range of $\varepsilon$. In Figure 2), the fast VLS-DSR light signals observably eliminate as a possibility for rest the state of motion associated with $\varepsilon = 3/4$.

Ultimately, using VLS-DSR signals of arbitrarily high energy and speed, an observer could effectively observe in multiple places at once, thereby skirting the synchronization dilemma and determining rest, to an accuracy limited only by clock precision, as Sjödin would have understood [11]. As Reichenbach noted [9b], finite but arbitrarily fast velocities, "...would suffice to define absolute simultaneity as a limit."

### IV. Observing the Ether, Using VLS-DSR Light of Any Two Different Speeds.

Suppose signals such as Poincaré described, for which $t''_1$, $t''_2$, and $t''_3$ are analogous to $t_1$, $t_2$, and $t_3$, with which we would have:

4) $$t''_3 - t''_1 \neq t_3 - t_1.$$

And suppose that our observer at clock $A$ sends two signals, one of such as we have just supposed and the other of ordinary light (with energy-independent speed), so that the two arrive at $B$ and are reflected back, simultaneously ($t''_2 = t_2$). With either the slower or faster signals ordinary light, Figure 3) shows that our observers could bring their clocks to rest, by adjusting their motion so that, along with $t''_2 = t_2$:

5) $$t_3 - t''_3 = t''_1 - t_1.$$

Figure 3)

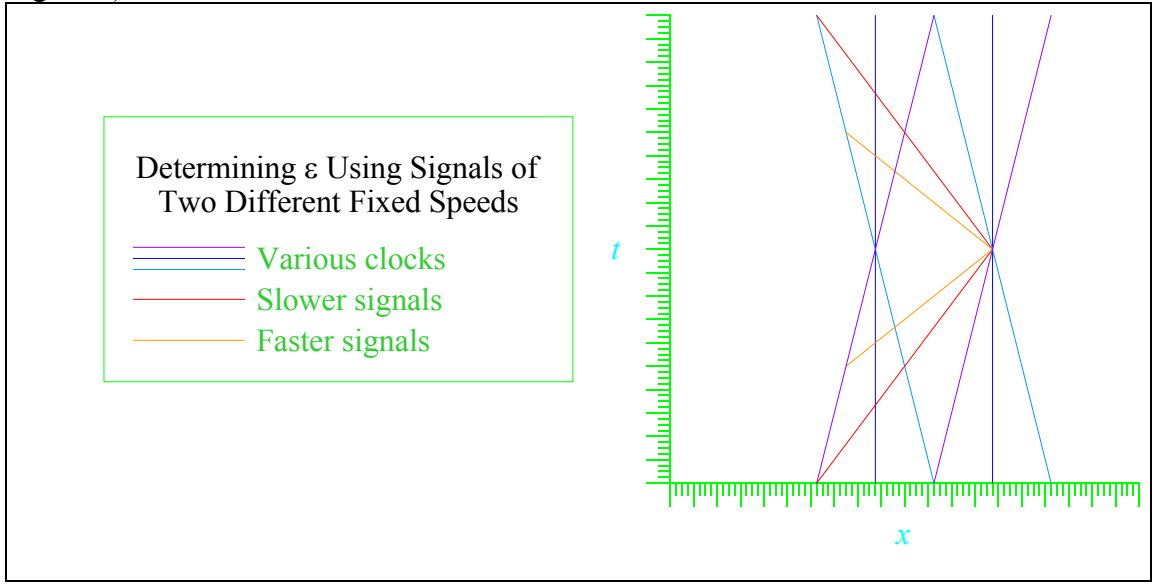

While VLS-DSR light of a particular energy would travel at a particular fixed speed, VLS-DSR light, emitted or reflected by a clock not at rest, would be Doppler shifted; so were our clocks not at rest, the rest-frame speed, of VLS-DSR light emitted from *A* to *B*, would not equal that of the corresponding light reflected from *B* to *A*. However, were the emitting purple clock, in Figure 3), to emit VLS-DSR light signals, the signals would be Doppler shifted up in energy, with the nonlinear variation of light speeds ensuring an interval between emission times greater than that in Figure 3), whether the light speeds increase or decrease with energy. The return signals would then be Doppler shifted down, with the interval between return times not equal to that between emission times. Were the emitting light-blue clock to emit VLS-DSR light signals, the signals would be Doppler shifted down in energy, with the interval between emission times smaller than that in Figure 3), and the interval between return times again not equal to that between emission times. Thus, the Doppler effect would be symmetric for clock motion about rest; so our observers could still bring their clocks to rest, using emitted VLS-DSR light of two different energies, by achieving analogies to $t''_2 = t_2$ and Equation 5).

## V. General Comments and Conclusions.

Aspects of VLS-DSR's high-energy character, not discussed above, cannot prevent the above results, since for each state of motion, the same stationary reference clocks make any measurements, regardless of any measured object's energy. Quantum mechanics has no impact, because VLS-DSR must apply at classical scales, with its deviating dispersion laws governing effectively classical objects propagating in effectively classical space-times. And even restriction of significant deviations in the dispersion laws, to the Planck scales at which both gravity and quantum effects are significant, holds no hope, especially in VLS-DSR's primary intended application [1] to photons arriving from cosmologically distant gamma-ray bursts, whereby, "In some cases the quantum gravity dispersion effect would predict these arrivals to be delayed or advanced by days to months...".

Experiments [1] may show speeds of light that vary with energy, possibly providing insights into quantum gravity but certainly empirically establishing the ether. Therefore, VLS-DSR obviates itself, since it answers the question posed by two of its most prominent proponents [5] (with $l_p$ and $E_p$ respectively the *Planck Length* and *Planck Energy*), "*In whose reference frame are $l_p$ and $E_p$ the thresholds for new phenomena?*"

VLS-DSR formalism conceivably could survive with an ether interpretation, possibly after modification; but with relativity gone, a new motivation would be necessary.

Our results apply to all theories with energy-dependent speeds of light, like those of gravity and cosmology that are associated with VLS-DSR (see [1,8] for reviews that include recent such theories, as well as various other theories to which our techniques would also apply).

## VI.  Acknowledgements.

Great thanks to Dr. Allen Janis, whose questions and comments improved this work.

## VII.  References.